# AdapSCA-PSO: An Adaptive Localization Algorithm with AI-Based Hybrid SCA-PSO for IoT WSNs

Ze Zhang
*School of Advanced Technology*
*Xi'an Jiaotong-Liverpool University*
Suzhou, China
Ze.Zhang20@alumni.xjtlu.edu.cn

Qian Dong*
*School of Advanced Technology*
*Xi'an Jiaotong-Liverpool University*
Suzhou, China
Qian.Dong@xjtlu.edu.cn

Wenhan Wang
*College of Environmental Science and Engineering*
*Ocean University of China*
Qingdao, China
wwh18345363561@163.com

*Abstract*—The accurate localization of sensor nodes is a fundamental requirement for the practical application of the Internet of Things (IoT). To enable robust localization across diverse environments, this paper proposes a hybrid meta-heuristic localization algorithm. Specifically, the algorithm integrates the Sine Cosine Algorithm (SCA), which is effective in global search, with Particle Swarm Optimization (PSO), which excels at local search. An adaptive switching module is introduced to dynamically select between the two algorithms. Furthermore, the initialization, fitness evaluation, and parameter settings of the algorithm have been specifically redesigned and optimized to address the characteristics of the node localization problem. Simulation results across varying numbers of sensor nodes demonstrate that, compared to standalone PSO and the unoptimized SCAPSO algorithm, the proposed method significantly reduces the number of required iterations and achieves an average localization error reduction of 84.97%.

## I. INTRODUCTION

Driven by the fast-paced advancement of the Internet of Things (IoT), the industrial IoT has increasingly become the foundational infrastructure of intelligent manufacturing. Wireless Sensor Networks (WSNs), consisting of numerous sensor nodes, enable collaborative operations in workshops, warehouses, and smart factories through real-time data collection and control[1]. These sensor nodes include a small number of anchor nodes with known locations and many unknown nodes whose positions must be estimated. Accurate localization is essential for the implementation and development of digital twin technologies and autonomous IoT systems, where location information serves as a cornerstone[2].

Due to environmental constraints, particularly in indoor scenarios, traditional GPS-based localization solutions are often infeasible[3]. In such contexts, meta-heuristic algorithms, a subset of artificial intelligence (AI) techniques, offer promising solutions to the node localization problem in WSNs. Compared to conventional mathematical approaches, meta-heuristic algorithms are more effective in handling nonlinear, multi-modal optimization problems. In contrast to deep learning models such as CNNs and Transformers, meta-heuristics are lightweight and easily deployable, making them a better fit for distributed WSNs applications.

To refine both the localization accuracy and convergence speed, this study introduces a node localization method based on the fusion of Sine Cosine Algorithm (SCA) alongside Particle Swarm Optimization (PSO) methods. The proposed AdapSCA-PSO algorithm integrates an adaptive control strategy. This strategy dynamically switches between the two algorithms to achieve superior convergence speed and localization accuracy. The overall workflow of AdapSCA-PSO algorithm is illustrated in Figure 1. The key contributions of this study can be outlined as:

1) **Adaptive hybrid SCA-PSO algorithm.** The proposed adaptive control strategy switches between SCA and PSO. SCA is effective in global search, while PSO excels at local search. This approach aims to achieve superior convergence speed and localization accuracy.

2) **Optimized initialization and fitness evaluation.** The algorithm initializes particles through connecting unknown nodes to the anchor nodes with the fewest hops. This strategy enhances both efficiency and stability without increasing computational complexity. Moreover, the fitness function focuses on minimizing the ranging error among neighbors.

3) **Adaptive design tailored for IoT applications.** The proposed method supports the distributed nature of WSNs deployments and achieves a well-balanced trade-off among localization accuracy, convergence speed, and robustness. This makes it particularly well-suited to meet the stringent spatial accuracy and real-time performance demands of industrial IoT.

The structure of this paper is as follows. Section II discusses the related background and introduces the key technologies employed in this study. Section III provides a description of the proposed AdapSCA-PSO. Section IV presents the simulation evaluation of the algorithm's localization performance. Finally, Section V provides the conclusion.

Corresponding author: Qian Dong. This work was supported by the XJTLU Research Development Fund (RDF) [grant No. 22-02-096]; Natural Science Foundation of Guangdong Basic and Applied Basic Research Fund-General Project [grant No. 2022A1515011309].

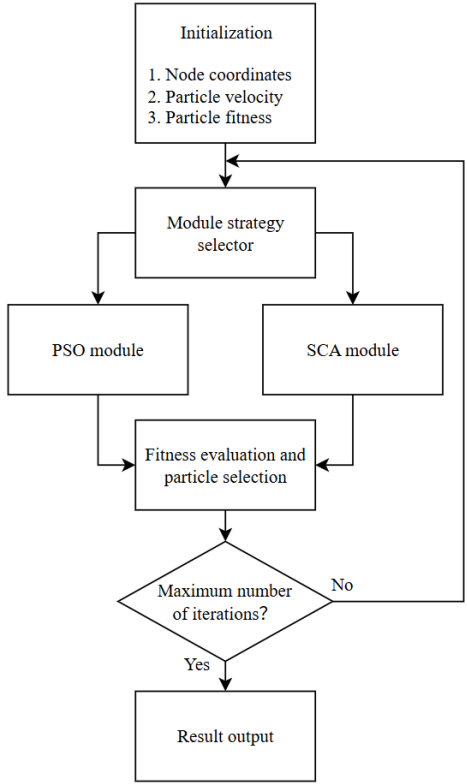

Figure 1. Main workflow of the AdapSCA-PSO algorithm.

## II. TECHNICAL BACKGROUND

This section reviews the meta-heuristic localization algorithms widely applied in WSNs. Particular emphasis is on the foundational algorithms of PSO and SCA.

### A. Meta-heuristic Localization Algorithm

In addition to PSO and SCA, several other mature meta-heuristic algorithms have been widely applied to WSNs localization, including Genetic Algorithm (GA), Ant Colony Optimization (ACO), and Artificial Bee Colony (ABC). These advanced optimization algorithms are typically inspired by collective behaviors in nature and are particularly well-suited for solving complex optimization problems. In the context of WSNs localization, the position of each unknown node is generally estimated by optimizing distance-based measurements between nodes, such as Received Signal Strength Indicator (RSSI) and Time of Arrival (ToA).

PSO is favored due to its simplicity and minimal parameter tuning, which contribute to the fast convergence. However, its global search capability is relatively limited. Conversely, algorithms such as SCA and GA exhibit stronger global search abilities, but their high degree of randomness can hinder the attainment of accurate localization results. ACO and ABC algorithms offer high localization accuracy. However, their algorithmic complexity and large number of control parameters place a heavy computational burden on sensor nodes, which often slows down convergence rates [4].

### B. Working Principle of PSO

In PSO, each particle signifies a possible candidate solution. At the initial stage, the particle swarm is typically initialized with random positions and velocities. Each particle then evaluates its fitness according to the objective function and adjusts its position[5]. The update rules for a particle's velocity and position are defined as:

$$v_i(t+1) = \omega \cdot v_i(t) + c_1 rand(p_i - x_i(t)) + c_2 rand(g - x_i(t)), \quad (1)$$

$$x_i(t+1) = x_i(t) + v_i(t+1), \quad (2)$$

where $v_i(t)$ and $x_i(t)$ denote the velocity and position of the $i$-th particle at iteration $t$. The parameter $p_i$ refers to the personal best position found by particle $i$, and $g$ represents the best position among all particles. The parameters $c_1$ and $c_2$ are acceleration coefficients that balance the influence of the personal best and global best positions. The parameter $\omega$ denotes the inertia weight, which helps maintain the momentum of the particle and ensures smooth movement through the search space, and the update formula is as:

$$\omega = \omega_{max} - \frac{\omega_{max} - \omega_{min}}{T} \cdot t, \quad (3)$$

where $\omega_{max}$ and $\omega_{min}$ are the upper and lower bounds of the inertia weight, and $T$ denotes the total number of iterations.

### C. Working Principle of SCA

In a typical SCA, the particle representation, initialization process, fitness evaluation, and convergence criteria are generally consistent with those in PSO. The key difference lies in the position update mechanism. Unlike PSO, which relies on both personal and global best positions, SCA focuses on guiding each particle toward the global optimum using a stochastic combination of sine and cosine functions along with several random coefficients. This design endows SCA with strong global search capabilities and a high degree of exploration[6]. The equation for updating the position of each particle is as:

$$x_i(t+1) = \begin{cases} x_i(t) + r_1 \sin(r_2) |r_3 g - x_i(t)|, r_4 < 0.5 \\ x_i(t) + r_1 \cos(r_2) |r_3 g - x_i(t)|, r_4 \geq 0.5 \end{cases}, \quad (4)$$

$$r_1(t) = a(1 - t/T). \quad (5)$$

The update process (4) involves four random coefficients $r_1$ to $r_4$. Specifically, $r_1 \in [0, a]$ controls the step size during the iteration to ensure convergence, where $a$ is the initial maximum amplitude, updated according to (5). The coefficient $r_2 \in [0, 2\pi]$ controls the oscillation frequency of sine and cosine functions. The coefficient $r_3 \in [-1, 1]$ controls the direction and amplitude of the particle's movement toward the global optimum. The coefficient $r_4 \in [0, 1]$ determines whether the sine or cosine function is applied during each iteration.

## III. METHODOLOGY

This section presents the proposed AdapSCA-PSO in three parts. Initially, the initialization process describes the initialization of particle velocities and positions. Second, an adaptive strategy selector is introduced, which dynamically switches between the SCA and PSO modules based on the current iteration number. Lastly, the procedure for evaluating the fitness values of particles is explained.

*A. Initialization*

In most meta-heuristic algorithms, the initial velocities and positions of particles are randomly assigned. However, such initialization often results in unpredictable velocities and irregular particle distributions. These irregularities can degrade both the accuracy and efficiency of the localization process.

To overcome these challenges, the proposed method exploits the spatial structure of nodes and the multi-hop communication patterns in WSNs. For each unknown node, the anchor node with the smallest number of hops away from it is selected. The initial position of the unknown node's particle is constrained within a one-hop radius of its nearest anchor node, thereby reducing the initial search space. Additionally, the initial velocity of the particle is scaled proportionally to the hop count between the unknown node and the nearest anchor node, enhancing search efficiency across different network topologies. The specific initialization formulas for the velocity and position of the particle $i$ are defined as:

$$v_i(0) = \delta \cdot h_i(2r - 1), \quad r \sim U(0,1), \tag{6}$$

$$x_i(0) \in \{x \in R^2 | \parallel x - a_z \parallel \leq L\}, \tag{7}$$

where $v_i(0)$ and $x_i(0)$ denote the initial velocity and position of the $i$-th particle. In (6), $\delta$ is an adjustable base velocity coefficient, $h_i$ is the hop count between the unknown node and its nearest anchor node, and $r \in [0, 1]$ is a uniformly distributed random number. In (7), $x$ denotes a candidate solution, and $R^2$ represents the two-dimensional localization space. The parameter $a_z$ represents the coordinates of the nearest anchor node, and $L$ is the maximum one-hop communication range.

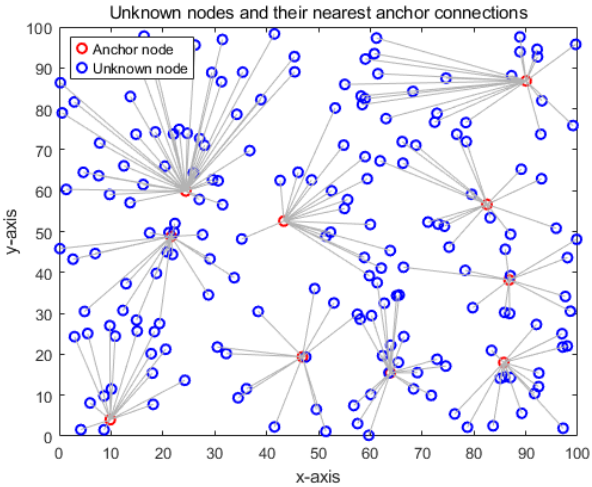

Figure 2. Initial node location estimation and approximate movement trajectories in the AdapSCA-PSO algorithm.

An illustrative example is provided in Figure 2. All unknown nodes are connected to their nearest anchor nodes. The distance of one hop around the position of the connected anchor node represents the constrained region for initializing particle positions of unknown nodes. The length of each gray line is positively correlated with the hop count and reflects the corresponding particle's initial maximum velocity. The gray lines also represent approximate particle movement trajectories during the iterative optimization process.

*B. Module Strategy Selector*

In meta-heuristic algorithms, achieving a balance between exploration and exploitation is crucial for ensuring efficiency and accuracy. To address this, the proposed AdapSCA-PSO algorithm combines SCA, which exhibits strong global search ability, with PSO, which is more effective in local search. The design intention is for the SCA module to dominate in the early iterations, allowing broad global exploration. As the process evolves, the PSO module becomes more active to refine the position and improve localization accuracy. To support this adaptive transition, an exponential probability-based module selector is employed. The selection mechanism is governed by the following update strategy:

$$x_i(t+1) = \begin{cases} \text{SCA}(x_i(t)), \text{if } s < e^{-\beta \cdot t/T} \\ \text{PSO}(x_i(t)), \text{otherwise} \end{cases}, \tag{8}$$

where $x_i(t)$ represents the position of the $i$-th particle at iteration $t$, with $i \in (1, i_{max})$. The parameter $s \sim U(0,1)$ is drawn from a uniform distribution used to determine the switching between the two modules, and $\beta$ is a control parameter that regulates the rate of transition. Therefore, the value of the expression $e^{-\beta \cdot t/T}$ ranges between $e^{-\beta}$ and 1. This strategy enables a smooth and adaptive shift from the SCA module to the PSO module over the course of the optimization process. By predominantly employing the SCA algorithm in the early phase and transitioning to the PSO algorithm in the later phase, the algorithm maintains sufficient diversity among particles, thereby enhancing the likelihood of identifying the global optimal solution.

*C. Fitness Evaluation*

In some traditional meta-heuristic localization algorithms, auxiliary localization methods such as DV-Hop are used to define the fitness function. However, such methods cause the effectiveness of the meta-heuristic algorithm to rely strongly on the precision of the auxiliary technique, while also increasing algorithmic complexity. To address this limitation, the proposed method directly leverages the inherent ranging capabilities of WSNs. Specifically, the fitness function is defined by minimizing the error between the measured distances and the estimated positions of neighboring nodes. The fitness function for node $j$ is formulated as:

$$f(X_j) = \sum_{k \in N_j} w_k (d_{jk} - \parallel X_j - X_k \parallel)^2, \tag{9}$$

where $N_j$ denotes the set of neighboring nodes within one-hop range of node $j$. The parameter $w_k$ reflects the type of node $k$, where $w_k = 0.8$ if node $k$ is an anchor node, and $w_k = 0.2$ if node $k$ is an unknown node. The term $d_{jk}$ denotes the distance between nodes $j$ and $k$. The parameters $X_j$ and $X_k$ represent the predicted locations of the nodes $j$ and $k$, and $\parallel X_j - X_k \parallel$ represents the Euclidean distance between nodes $j$ and $k$.

## IV. NUMERICAL RESULTS

An evaluation of the AdapSCA-PSO algorithm's localization accuracy is conducted in this section under four varied localization scenarios. The calculated localization errors are compared against those of the DV-Hop algorithm, the standard PSO algorithm with default parameters, and the unoptimized SCAPSO algorithm.

### A. Simulation Setups

The simulation experiments are conducted within a $100 \times 100$ m$^2$ area, considering two different node scenarios: 100 nodes and 200 nodes. To better reflect real-world deployment conditions, when the node number is 100, the maximum communication distance $L$ is set to 30 meters. When the node number is 200, $L$ is reduced to 15 meters. For both node configurations, two anchor node ratios are considered, 10 percent and 20 percent, in order to examine the impact of anchor node density on the localization accuracy of the four methods. Each of the four scenario settings is simulated 50 times to ensure statistical robustness. Table 1 outlines the key parameter values used to implement the proposed algorithm.

Table 1. Algorithm parameter settings.

| Symbol | Description | Value(s) |
| --- | --- | --- |
| $\omega_{max}$ | Upper bound of the inertia weight. | 0.9 |
| $\omega_{min}$ | Lower bound of the inertia weight. | 0.4 |
| $i_{max}$ | Number of particles. | 30 |
| $c_1$ | Acceleration coefficient of personal optimal solution. | 2.2 |
| $c_2$ | Acceleration coefficient of global optimal solution. | 1.8 |
| $T$ | Maximum number of iterations. | 60 |
| $a$ | Initial maximum amplitude. | 2.5 |
| $\delta$ | Adjustable base velocity coefficient. | 0.5 |
| $L$ | Maximum one-hop communication range. | 15, 30 |
| $\beta$ | Control parameters for adjusting the conversion rate. | 3 |

The hyperparameter settings used in the algorithm are based on prior research and empirical experience as reported in [5] and [6]. Notably, the values of $c_1$ and $c_2$ are adjusted to enhance the local exploitation capability of the PSO component, thereby improving the overall positioning accuracy. Additionally, the parameter $a$ is slightly increased to facilitate faster global exploration during the early iterations, which contributes to quicker convergence and improved computational efficiency.

### B. Assessment Criteria

To assess the accuracy of localization, the distance between the true and estimated positions of all unknown nodes is calculated for each experiment as the primary performance indicator. The average localization error is computed using the following formula:

$$AvgError = \frac{\sum_{e=1}^{UN} \|X_e^{ture} - X_e^{est}\|}{UN}, \quad (10)$$

where $AvgError$ denotes the average coordinate error for all unknown nodes in a single simulation experiment, $UN$ is the total number of unknown nodes, and $\|X_e^{ture} - X_e^{est}\|$ is the distance between the true and estimated positions of unknown node $e$. All four algorithms are evaluated across 50 independent simulations under four different localization scenarios.

### C. Evaluation Results of AdapSCA-PSO

Based on the assessment criteria and localization scenarios for unknown nodes, Figure 3 depicts the variation in average localization errors across the four methods. Correspondingly, Table 2 presents the numerical results of the average localization errors for each method under different localization scenarios.

Table 2. Average localization error of four algorithms in different scenarios.

| Scenarios | DV-Hop | PSO | SCAPSO | AdapSCA-PSO |
| --- | --- | --- | --- | --- |
| Scenario 1 | 10.9359m | 8.8107m | 8.6088m | 0.6722m |
| Scenario 2 | 9.2539m | 6.4227m | 6.4436m | 0.4778m |
| Scenario 3 | 8.3332m | 5.8671m | 5.7791m | 1.4689m |
| Scenario 4 | 7.6721m | 4.5913m | 4.5987m | 0.9194m |

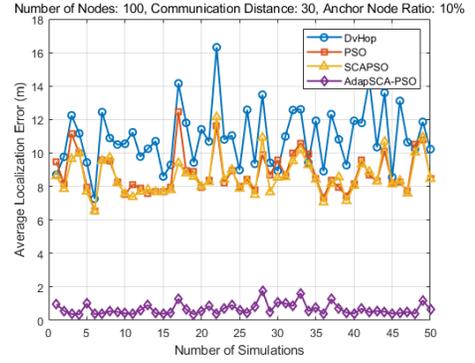

(a) Scenario 1: Small network with few anchors.

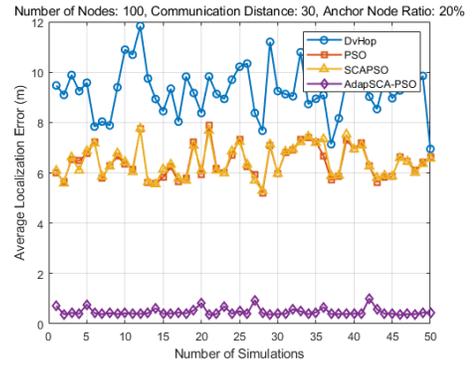

(b) Scenario 2: Small network with more anchors.

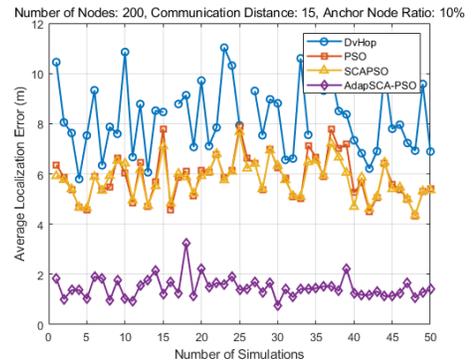

(c) Scenario 3: Large network with few anchors.

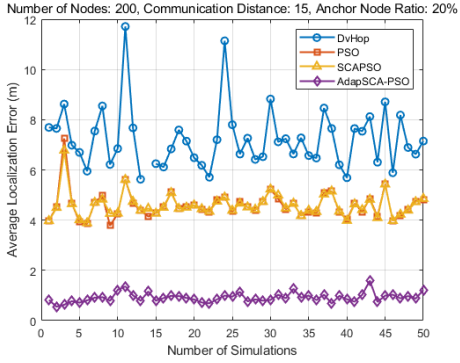

(d) Scenario 4: Large network with more anchors.

Figure 3. Comparison of average localization error across different scenarios.

As illustrated in Figure 3, the DV-Hop algorithm consistently exhibits the highest localization error among the four methods. Although the fitness functions of traditional PSO and SCAPSO rely on DV-Hop, their iterative optimization processes allow them to account for the overall localization scenario and converge toward global optima. Consequently, both algorithms achieve lower and more stable errors compared to DV-Hop. AdapSCA-PSO algorithm outperforms all three benchmarks, maintaining the lowest average error and highest stability across all scenarios.

Analyzing the four deployment scenarios, an increasing number of anchor nodes improves the localization accuracy for all algorithms. This is because DV-Hop inherently depends on anchor node information, and the fitness functions of PSO and SCAPSO are directly influenced by DV-Hop's accuracy. The AdapSCA-PSO algorithm, on the other hand, benefits from additional anchor nodes by obtaining more reliable one-hop neighbors during the fitness evaluation process. An increase in the number of anchor nodes significantly enhances localization accuracy, with an average error reduction of 33.17%.

When the total number of nodes increases and the maximum communication range decreases, the localization errors of the three benchmark algorithms also decline. This improvement is attributed to the higher node density, which results in more direct communication paths between unknown nodes and anchor nodes, thereby enhancing distance estimation accuracy. For AdapSCA-PSO, although its fitness evaluation relies solely on one-hop neighbors, the reduction in communication distance causes a slight decrease in the number of effective neighbors, resulting in a minor rise in localization error. However, the error remains significantly lower than those of the other three algorithms.

Figure 4 presents the convergence curves of PSO, SCAPSO, and the proposed AdapSCA-PSO algorithm. AdapSCA-PSO begins with a lower initial fitness value due to its WSN-specific initialization strategy. In the subsequent iterations, the SCA module demonstrates strong global convergence capabilities, achieving lower fitness values more rapidly than PSO. With its adaptive switching mechanism and optimized parameter settings, AdapSCA-PSO exhibits faster convergence speed and ultimately achieves the lowest final fitness.

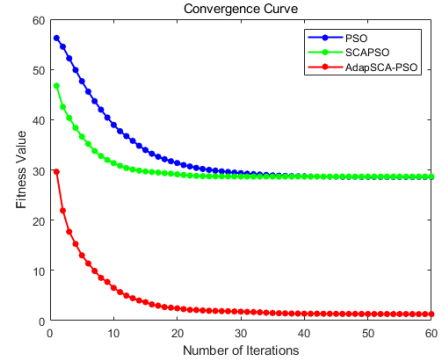

Figure 4. Convergence curves of three meta-heuristic algorithms.

In conclusion, the proposed AdapSCA-PSO algorithm demonstrates improved results in localization accuracy and convergence stability. In comparison to DV-Hop, PSO, and SCAPSO, the proposed AdapSCA-PSO reduces the average localization error by 89.77%, 84.97%, and 84.84% across four different localization scenarios.

## V. CONCLUSION

This paper presents AdapSCA-PSO, a hybrid adaptive meta-heuristic algorithm specifically designed to mitigate localization inaccuracies in WSNs for IoT applications. The design intention is for the SCA module to dominate in the early iterations, allowing broad global exploration. As the process evolves, the PSO module becomes more active to refine the position and improve localization accuracy. Through an adaptive switching strategy, the proposed method achieves more accurate node localization. Compared with PSO and unoptimized SCAPSO, the proposed AdapSCA-PSO improves localization accuracy by 84.97%. Overall, its high accuracy and robustness demonstrate strong potential for practical deployment for localization in WSN-based IoT systems.